\documentclass[runningheads]{llncs}
\usepackage{graphicx}
\usepackage{color}
\usepackage[list=true]{subcaption}
\usepackage{enumerate}
\usepackage{multicol}
\usepackage{rotating}
\usepackage{mathtools}
\usepackage{sidecap}
\usepackage{siunitx}
\usepackage{placeins}
\usepackage{comment}
\usepackage{sidecap}

 \usepackage{xcolor}

\usepackage{algpseudocode}
\usepackage{algorithm}
\usepackage{multirow}
\usepackage{makecell}
\usepackage{paralist}
\usepackage{cancel}
\usepackage{color, colortbl}
\usepackage[labelfont=bf,textfont=it,belowskip=0pt,aboveskip=5pt,tableposition=top]{caption}
\usepackage[colorlinks=true,
citecolor=blue,
filecolor=black,
linkcolor=blue,
urlcolor=blue]{hyperref}
\usepackage{longtable}
\usepackage{graphbox}
\usepackage{xspace}

\graphicspath{{figures/}{tables/}}

\newcounter{runidnum}

\newcommand{\secref}[1]{\S\ref{#1}}

\newcolumntype{R}{>{\columncolor{gray!20}}r}
\newcolumntype{L}{>{\columncolor{gray!20}}l}
\definecolor{Gray}{gray}{0.85}
\newcolumntype{a}{>{\columncolor{Gray}}c}

\newcommand{\elltwoT}{\ensuremath{\mu_{T,L_2}}} 

\newcommand{\mpPatientBrain}[1][]{\ifthenelse { \equal {#1} {} }
    { \vect{m}_D }   
    { \vect{m}_{D,#1} } }  

\newcommand{\mpWarpedAtlasBrain}[1][]{\ifthenelse { \equal {#1} {} }
    { \vect{m}_A^{(1,1)} }   
    { \vect{m}_{A,#1}^{(1,1)} }}   

\newcommand{\ns}[1]{\ensuremath{\mathbf{#1}}}

\newcommand{\idiv}{\ensuremath{\mbox{div}\,}}
\newcommand{\igrad}{\ensuremath{\nabla}}

\newcommand{\p} {\partial}

\newcommand{\vect}[1]{\boldsymbol{#1}} 
\newcommand{\mat}[1]{\boldsymbol{#1}}  

\newcommand{\bipa}{\begin{inparaenum}[(\itshape i\upshape)]}
\newcommand{\eipa}{\end{inparaenum}}

\newcommand{\bipasub}{\begin{inparaenum}[(\itshape a\upshape)]}
\newcommand{\eipasub}{\end{inparaenum}}

\newcommand{\defeq}{\ensuremath{\mathrel{\mathop:}=}}

\newcommand{\T}{\ensuremath{\mathsf{T}}}

\definecolor{sred}{cmyk}{0.01,0.98,0,0.2} 
\reversemarginpar
\newcommand{\mmargin}[1]{{\marginpar{\em\tiny #1}}}\renewcommand{\mmargin}[1]{}

\newcommand{\beginsupplement}{%
    \setcounter{table}{0}
    \renewcommand{\thetable}{S\arabic{table}}%
    \setcounter{figure}{0}
    \renewcommand{\thefigure}{S\arabic{figure}}%
}
\begin{document}
\title{Multiatlas Calibration of Biophysical Brain Tumor Growth Models with Mass Effect} 
\titlerunning{Multiatlas Calibration of Tumor Growth Models with Mass Effect}
%
\author{Shashank Subramanian \inst{1} \and
Klaudius Scheufele\inst{1} \and
Naveen Himthani\inst{1} \and
George Biros\inst{1}}
\authorrunning{S. Subramanian et al.}
%
\institute{Oden Institute, University of Texas at Austin,
    201 E. 24th Street, Austin, Texas, USA \\
\email{\{shashank,naveen,biros\}@oden.utexas.edu}, \email{kscheufele@austin.utexas.edu}}

\maketitle              
\begin{abstract}
  We present a \emph{3D fully-automatic} method for the calibration of partial differential equation (PDE) models of glioblastoma (GBM) growth  with \emph{``mass effect''}, the deformation of brain tissue due to the tumor. We quantify  the mass effect, tumor \emph{proliferation},  tumor \emph{migration}, and the localized \emph{tumor initial condition} from a \emph{single} multiparameteric Magnetic Resonance Imaging (mpMRI) patient scan. The PDE is a reaction-advection-diffusion partial differential equation  coupled with linear elasticity equations to capture mass effect. The single-scan calibration model is notoriously difficult because the precancerous (healthy) brain anatomy is unknown.   To solve this inherently ill-posed and ill-conditioned optimization problem, we introduce a novel inversion scheme that uses \emph{multiple brain atlases} as proxies for the healthy precancer patient brain resulting in robust and reliable parameter estimation.
We apply our method on both synthetic and clinical datasets representative of the heterogeneous spatial landscape typically observed in glioblastomas to demonstrate the validity and performance of our methods. In the synthetic data,  we report calibration errors (due to the ill-posedness and our solution scheme) in the 10\%-20\% range. In the clinical data, we report good quantitative agreement with the observed tumor and qualitative agreement with the  mass effect (for which we do not have a ground truth). Our method uses a minimal set of parameters and  provides both global and local quantitative measures of tumor infiltration and mass effect.
\keywords{Glioblastoma  \and Mass effect \and Tumor growth models \and Inverse problems.}
\end{abstract}

\section{Introduction}
\label{sec:intro}
Gliomas are the most common primary brain tumors in adults. Glioblastomas are high grade gliomas with poor prognosis.  A significant challenge in the characterization of these tumors involves mass effect (the deformation of the surrounding healthy tissue due to tumor growth) along with biomarkers representing the tumor aggressiveness and growth dynamics. The integration of mathematical models with clinical imaging data holds the enormous promise of robust,  minimal, and explainable models that  quantify cancer growth and connect cell-scale phenomena to organ-scale, personalized, clinical observables~\cite{rockne-e19,yankeelov-miga13,Swanson:2008:survivalResection}.
Here, we focus on the calibration of mathematical tumor growth models with clinical imaging data from a \emph{single} pretreatment scan in order to assist in diagnosis and treatment planning. Longitudinal pretreatment scans are rare for GBMs since most patients seek immediate treatment. For this reason, we need algorithms that rely only on one mpMRI scan.

\textit{Contributions:} The single-scan calibration problem is formidable for two main reasons: the tumor initial condition (IC) and the subject's healthy precancer anatomy are unknown. Using brain anatomy symmetry does not apply to all subjects so typically an atlas is used as a proxy to the healthy subject brain. However, natural anatomical differences between the atlas and the subject interfere with tumor-related deformations; disentangling the two is hard.  In light of these difficulties, our contributions are as follows:
\textbullet \,  Based on the method described in~\cite{Subramanian20IP}, we propose a novel multistage scheme for inversion: first we estimate the tumor initial conditions, then, given an atlas, we invert for the three scalar model parameters (proliferation, migration, mass effect). We repeat this step for several atlases and we compute expectations of the observables (see~\secref{sec:methods} and~\secref{sec:scheme}).  These calculations are quite expensive. However, the entire method runs in parallel on GPUs so that 3D inversion in $128^3$ resolution takes less than an hour and $256^3$ resolution takes about six hours.
\textbullet \, We use synthetic data (for which we know the ground truth) in order to estimate the errors associated with our numerical scheme and we validate our method on a set of clinical mpMRI scans. We report these results in~\secref{sec:results}. 

\textit{Related work:} The most common mathematical models for tumor growth dynamics are based on reaction-diffusion PDEs~\cite{murray1989mathematical,swanson2000quantitative,konukoglu2010image,mang2012biophysical,swanson2002virtual}, which have been coupled with mechanical models to capture mass effect~\cite{hogea2007modeling,clatz2005realistic,hormuth2018,Subramanian:2019a}. While there have been many studies to calibrate these models using inverse problems~\cite{konukoglu2010image,Colin:2014a,Hogea:2008b,Gholami:2016a,Knopoff:2013a,mang2012biophysical,Gooya:2012a,Scheufele:2017a}, most do not invert for all unknown parameters (tumor initial condition and model parameters) or they assume the presence of multiple imaging scans (both these scenarios make the inverse problem more tractable). In~\cite{Subramanian20IP}, the authors presented a methodology to invert for tumor initial condition (IC) and cell proliferation and migration from a single scan; but their forward (growth) model does not account for mass effect. They demonstrate the importance of using a \emph{sparse tumor IC} (see \secref{sec:methods}) to correctly reconstruct for other tumor parameters. In \cite{Gholami2016-thesis}, the author considers mathematical aspects of inverting simultaneously for the tumor IC, tumor parameters, and mass effect but assumes known precancer brain anatomy.
Other than those works,  the current state of the art for single-scan biophysically-based tumor characterization is GLISTR~\cite{Gooya:2012a}.

The main shortcoming of GLISTR is that it requires manual seeding for the tumor IC and uses  a single atlas. Further, GLISTR uses deformable registration for both anatomical variations and mass effect deformations but does not decouple them (this is extremely ill-posed), since the primary goal of GLISTR lies in image segmentation. Finally, the authors in~\cite{Abler2019} present a 2D synthetic study to quantify mass effect. However, they assume known tumor IC and precancer brain.
To the best of our knowledge, we are not aware of any other framework that can fully-automatically calibrate tumor growth models with mass effect for all unknown parameters in 3D.
Furthermore, our solvers employ efficient parallel algorithms and GPU acceleration.

\section{Methods}
\label{sec:methods}
Before we describe our methods, we introduce the following notations: $c = c(\vect{x},t)$ is the tumor concentration ($\vect{x}$ is a voxel, $t$ is time) with \emph{observed} tumor data  $c_1$ and  \emph{unknown} tumor IC $c_0$; $\vect{m}(\vect{x},t) \defeq (m_{\text{WM}}(\vect{x},t), m_{\text{GM}}(\vect{x},t), m_{\text{CSF}}(\vect{x},t))$ is the brain segmentation into white matter (WM), gray matter (GM), and cerebrospinal fluid (CSF) with \emph{observed} pretreatment segmented brain $\vect{m}_1$ and \emph{unknown} precancer healthy brain $\vect{m}_0$; Additionally, ($\kappa$, $\rho$, $\gamma$) are scalars that represent the \emph{unknown} migration rate,  proliferation rate, and a mass effect parameter, respectively.

\medskip  \noindent
\textbf{Tumor growth mathematical model: }
Following~\cite{Subramanian:2019a}, we use a non-linear reaction-advection-diffusion PDE (on an Eulerian framework):
\begin{subequations}
    \label{e:fwd:mass}
    \begin{align}
    \p_t c + \idiv (c\vect{v}) - \kappa \mathcal{D} c  - \rho \mathcal{R} c & = 0 , ~~c(0) = c_0             && \mbox{in}~\Omega \times [0,1] \label{e:reac_diff}  \\
        \p_t \vect{m} + \idiv (\vect{m} \otimes\vect{v}) & = 0, ~~\vect{m}(0) = \vect{m}_0&&\mbox{in}~ \Omega \times [0,1] \label{e:m_transport}  \\
    \idiv (\lambda \igrad \vect{u} + \mu (\igrad \vect{u} + \igrad\vect{u}^\T)) & = \gamma \igrad c      && \mbox{in}~\Omega \times [0,1] \label{e:linear_elasticity} \\
    \p_t \vect{u} & = \vect{v}, ~~\vect{u}(0) = \vec{0}     \label{e:velocity}          && \mbox{in}~\Omega \times [0,1],
    \end{align}
\end{subequations}
where $\mathcal{D} :=  \idiv m_{\text{WM}} \igrad c$ is a diffusion operator; $\mathcal{R} := m_{\text{WM}} c(1 - c)$ is a logistic growth operator; 
Eq.~\eqref{e:reac_diff} is coupled to a linear elasticity equation (Eq.~\eqref{e:linear_elasticity}) with forcing $ \gamma \igrad c$, which is coupled back through a convective term with velocity $\vect{v}(\vect{x},t)$ which parameterizes the displacement $\vect{u}(\vect{x},t)$. 
The linear elasticity
model is parameterized by  Lam\`e coefficients $\lambda(\vect{x,t})$ and $\mu(\vect{x},t)$.
We note here that $\gamma = 0$ implies $\vect{u} = 0$, which reduces the tumor growth model to a simple reaction-diffusion PDE with no mass effect (i.e, Eq.~\eqref{e:reac_diff} with $\vect{v} = \vec{0}$).

Following~\cite{Subramanian20IP}, we parameterize $c_0 =  \boldsymbol{\phi}^T(\vect{x}) \vect{p} = \sum_{i=1}^{m} \phi_i(\vect{x}) p_i$;
where $\vect{p}$ is an $m$-dimensional parameterization vector, $\phi_i(\vect{x}) = \phi_i(\vect{x}-\vect{x_i}, \sigma)$ is a Gaussian function centered at point $\vect{x}_i$ with standard deviation $\sigma$, and $\boldsymbol{\phi}(\vect{x}) = \{\phi_i(\vect{x}) \}_{i = 1}^m$. Here, $\vect{x}_i$ are voxels that are segmented as tumor and $\sigma$ is one voxel, meaning $m$ can be quite large ($\sim$1000). This parameterization alleviates some of the ill-posedness associated with the inverse problem~\cite{Subramanian20IP}.

\medskip \noindent
\textbf{Inverse problem: }The unknowns in our growth model are $(\vect{p},\vect{m}_0,\kappa, \rho, \gamma)$. Our model is calibrated for these parameters using imaging data. We introduce an approximation to $\vect{m}_0$ (discussed in the numerical scheme), and estimate the rest through the following inverse problem formulation:
 \begin{gather}\label{eq:tumor_objective}
\min_{\vect{p},\kappa, \rho, \gamma}~ \mathcal{J} (\vect{p},\kappa, \rho, \gamma) :=
\frac{1}{2} \| \mat{O}c(1) - c_1 \|_{L_2(\Omega)}^2 + \frac{\beta}{2} \| \boldsymbol{\phi}^T \vect{p} \|_{L_2(\Omega)}^2
\end{gather}
\noindent subject to the reaction-advection-diffusion forward (growth) model $\mathcal{F}(\vect{p},\kappa, \rho, \gamma)$ given by Eq.~\eqref{e:fwd:mass}. Recall that $\kappa, \rho$, and $\gamma$ are scalars, but $\vect{p} \in\ns{R}^{m}$. 
The objective function minimizes the $L_2$ mismatch between the simulated tumor $c(1)$ at $t = 1$ and data $c_1$ and is balanced by a regularization term on the inverted initial condition (IC). $\mat{O}$ is an observation operator that defines the clearly observable tumor margin (see~\cite{scheufele2020automatic} for details).
Following~\cite{Subramanian20IP}, we further introduce the following constraints to our optimization problem: $\| \vect{p} \|_0 \leq s$ and $\max (\boldsymbol{\phi}^T \vect{p}) = 1$.
The first constraint restricts the tumor initial condition to a few Gaussians while the second enforces the assumption that $t = 0$ corresponds to the first time the tumor concentration reaches one at some voxel in the domain. Note that these two constraints are \emph{modeling assumptions}. They are introduced to alleviate the severe ill-posedness of the backward tumor growth PDE (see~\cite{Subramanian20IP} for details).

\subsection{Summary of our multistage inversion method}\label{sec:scheme}
To solve our inverse problem, we propose the following numerical scheme:
\begin{enumerate}[(\textbf{S}.1)]
    \item We solve Eq.~\eqref{eq:tumor_objective} using the \emph{simple} growth model with no mass effect $\mathcal{F}(\vect{p},\kappa, \rho, \gamma = 0)$\footnote{$\gamma = 0$ indicates no  mass effect and Eq.~\eqref{e:m_transport}, Eq.~\eqref{e:linear_elasticity} and Eq.~\eqref{e:velocity} are not needed.} for $(\vect{p}, \kappa, \rho)$. In this step, our \emph{precancer scan}, i.e. $\vect{m}_0$, is approximated as the patient brain with tumor regions replaced by white matter. In order to solve the resulting non-linear inverse problem, we employ the fast \emph{adjoint-based} algorithm outlined in~\cite{Subramanian20IP}.
    \item Next, due to mass effect, we need an estimate for $\vect{m}_0$. For this, we use $m_0(\vect{x}) = a(\vect{x})$, where $a(\vect{x})$ is an atlas or a scan from another healthy individual. We note that the \emph{inverted} tumor initial condition $\vect{p^\text{rec}}$ lies in the precancer scan space from (\textbf{S}.1). In order to make sure that the $\vect{p^\text{rec}}$ locations do not fall in anatomical structures such as ventricles (where the tumor does not grow), we register the patient\footnote{We simply mask the tumor region of the patient for this registration.} and the atlas (our new precancer scan), and transfer (warp) the initial condition to the atlas with the deformation map. This registration is unnecessary if the precancer scan is known.
    \item Finally, we solve Eq.~\eqref{eq:tumor_objective} constrained by $\mathcal{F}(\vect{p} = \vect{p^\text{rec}},\kappa, \rho, \gamma)$ for $(\kappa, \rho, \gamma)$. Since we have only three unknown parameters, we use first order finite differences to approximate the gradient of the objective.
\end{enumerate}

We repeat (\textbf{S}.2) and (\textbf{S}.3) for different atlases to make our inversion scheme less sensitive to the atlas selected. For the $\ell_2$ regularized solve in (\textbf{S}.1) and the inverse solve in (\textbf{S}.3), we use a quasi-Newton optimization method (L-BFGS) globalized by Armijo linesearch with gradient-based convergence criteria. For the registration in (\textbf{S}.2), we use the registration solver CLAIRE~\cite{Mang2018b}.

\medskip \noindent
\textit{Solver timings: } On average, the full multistage inversion on $128^3$ takes less than an hour using GPUs. The inversion in the different atlases are embarrassingly parallel. For $256^3$ (the resolution of our results), the full inversion takes an average of six hours. Finally, our optimization solvers converge in an average of 20-30 quasi-Newton iterations without any failures.

\section{Results}
\label{sec:results}
%
We ask the following four questions: \begin{enumerate}[\bf(Q1)\upshape] \item Given $\vect{p}$ (tumor IC) and $\vect{m}_0$ (precancer scan), can we reconstruct $(\kappa, \rho, \gamma)$ using (\textbf{S}.3) from~\secref{sec:scheme}? \item Given $\vect{m}_0$ but unknown $\vect{p}$, can we reconstruct $(\vect{p}, \kappa, \rho, \gamma)$ using (\textbf{S}.1) and (\textbf{S}.3)? \item With $\vect{m}_0$ and $\vect{p}$ unknown, can we reconstruct $(\vect{p}, \kappa, \rho, \gamma)$ using (\textbf{S}.1)-(\textbf{S}.3) taking different atlases as $\vect{m}_0$? \item How does our scheme perform on clinical data?\end{enumerate} 

 We use \emph{synthetic} data to answer (i)-(iii) and quantify our errors. For \emph{clinical} data, since the ground truth is unknown (we do not have longitudinal data), we evaluate our scheme qualitatively.

\begin{figure}[htbp!]
    \centering
    \includegraphics[width=\textwidth]{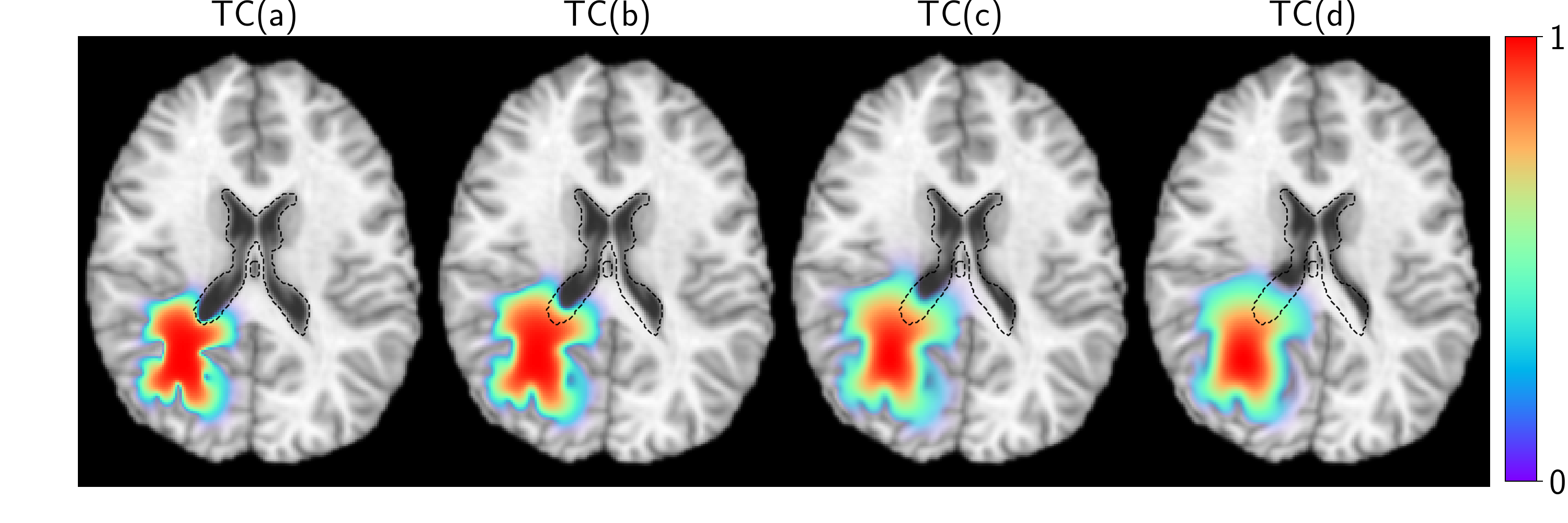}
    \caption{Synthetic patient T1 MRIs generated with Eq.~\eqref{e:fwd:mass}. The normalized tumor concentration is overlaid (color) along with the undeformed ventricles (black dashed contour) to indicate the variable extent of mass effect. } \label{fig:syn}
\end{figure}

\medskip \noindent
\textbf{(Q1) Known $\vect{m}_0$, known $\vect{p}$: }In this experiment, we generate data by growing synthetic tumors resembling clinical observations in a healthy atlas using different ground truth parameter combinations. The test-cases are aimed at simulating similar tumor volumes, but with varying amount of mass effect.  We consider the following variations for our parameter configurations:

\medskip
\begin{tabular}{ccclll}
    \textit{(i)}   &\textit{TC(a): } & no mass effect & $\gamma^\star = 0$  & $\rho^{\star}=12$ & $\kappa^{\star}=0.025$   \\
    \textit{(ii)}  &\textit{TC(b): } & mild mass effect & $\gamma^\star = 0.4$  & $\rho^{\star}=12$ & $\kappa^{\star}=0.025$   \\
    \textit{(iii)}  &\textit{TC(c): } & moderate mass effect & $\gamma^\star = 0.8$  & $\rho^{\star}=10$ & $\kappa^{\star}=0.05$   \\
    \textit{(iv)}  &\textit{TC(d): } & large mass effect & $\gamma^\star = 1.2$  & $\rho^{\star}=10$ & $\kappa^{\star}=0.025$,  \\
\end{tabular} \\

\noindent where $^\star$ represents the non-dimensionalized ground truth parameters. The tumors along with the deformed atlas are visualized in Fig.~\ref{fig:syn}. 
We report our inversion results in Tab.~\ref{tab:syn} (``True IC") with tumor initial condition and precancer scan taken as the ground truth. We report the relative errors in reconstructing parameters $\iota = \{\kappa, \rho, \gamma\}$ (if the ground truth is zero, then the error is absolute; \textit{TC(a)}), relative error in the two-norm of the displacement norm $u$, i.e., $\|\vect{u}\|_2$ (this field informs us of the extent of mass effect), relative error in the final tumor reconstruction, and the norm of the gradient to indicate convergence. We observe excellent reconstruction with relative errors less than 2\%.

 \medskip \noindent
\textbf{(Q2) Known $\vect{m}_0$, unknown $\vect{p}$: }We use our inversion scheme outlined in~\secref{sec:scheme}, where we first invert for the tumor initial condition using (\textbf{S}.1) and then the model parameters with \emph{known} precancer scan. We report our inversion results in Tab.~\ref{tab:syn} under ``Inverted IC". As expected, the errors increase due to the fact that we reconstructed $\vect{p}$ (IC) using the no-mass-effect model (images of the reconstructions included in {supplementary} Fig. S1). But we can still recover the model parameters quite well: the mass effect (indicated by the error in two-norm of the displacement norm $e_u$) is captured with relative errors less that 4\%; the reaction and diffusion coefficient also have good estimates of around 15\% and 22\% average relative errors respectively. Hence, our scheme exhibits good reconstruction performance.

\setlength\tabcolsep{6pt}
\begin{table}[h!]
    \caption{Inversion results assuming \textbf{known} $\vect{m}_0$, the anatomy of the subject before tumor occurrence. $u^\star$ is the ground truth displacement norm in number of voxels (1 voxel $\approx$ 0.9 mm), $e_\iota$ is the relative error for parameter $\iota$, $e_u$ is the relative error in the two-norm of the displacement field norm, $\|\vect{g}\|_2$ is the norm of the final gradient, and $\elltwoT$ is the relative error in final tumor concentration.  We report the tumor initial condition (IC) used in ``Tumor IC" -- True IC indicates synthetic ground truth IC and Inverted IC indicates our reconstruction using (\textbf{S}.1) and (\textbf{S}.3).\label{tab:syn}
    }
    \centering
    \makebox[\textwidth]{\centering
        \begin{tabular}{cc|aaa|aa|cc}
            \hline
            Test-case   & Tumor IC & $e_\gamma$ & $e_\rho$ & $e_\kappa$  & $\|u^\star\|_\infty$ & $e_u$ & $\|\vect{g}\|_2$ & $\elltwoT$  \\
            \hline    
            {\textit{TC (a)}}& True IC & \num{4.442900e-06} & \num{2.258333e-03} & \num{2.756000e-03} & 0 & \num{4.628200e-04} & \num{1.542016e-01} & \num{3.277710e-03} \\
            \textit{TC (b)} & True IC & \num{1.500000e-03} & \num{2.691667e-03} & \num{1.391600e-02} & 6.1& \num{7.270749e-04} & \num{5.697918e-04} & \num{3.516510e-03} \\
            \textit{TC (c)} & True IC & \num{1.322500e-03} & \num{2.500000e-05} & \num{2.140000e-04} &10 & \num{1.485759e-03} & \num{4.813955e-05} & \num{7.179340e-04} \\
            \textit{TC (d)} & True IC & \num{2.166667e-04} & \num{1.000000e-05} & \num{3.120000e-04} & 14.9 & \num{4.609798e-04} & \num{6.263090e-05} & \num{3.347420e-04}   \\        
            
            \hline
            \textit{TC (a)} & Inverted IC  & \num{2.946220e-03} & \num{1.632167e-01} & \num{2.082520e-01} & 0 & \num{2.856330e-01} & \num{1.268029e-02} & \num{2.422100e-01} \\
            \textit{TC (b)} & Inverted IC& \num{5.288500e-02} & \num{1.291250e-01} & \num{1.937200e-01} & 6.1 & \num{4.931780e-03} & \num{6.428073e-04} & \num{2.115820e-01}\\
            \textit{TC (c)} & Inverted IC&\num{3.536875e-02} & \num{1.385320e-01} & \num{2.552600e-01} & 10  & \num{1.458284e-02} & \num{3.540793e-04} & \num{1.654940e-01} \\
            \textit{TC (d)} & Inverted IC& \num{1.462500e-02} & \num{1.586730e-01} & \num{2.163040e-01} & 14.9 & \num{3.782162e-02} & \num{7.475129e-04} & \num{1.877340e-01}\\  
            
            \hline
            
    \end{tabular}}
\end{table}

\begin{figure}[h!]
    \centering
    \includegraphics[width=\textwidth]{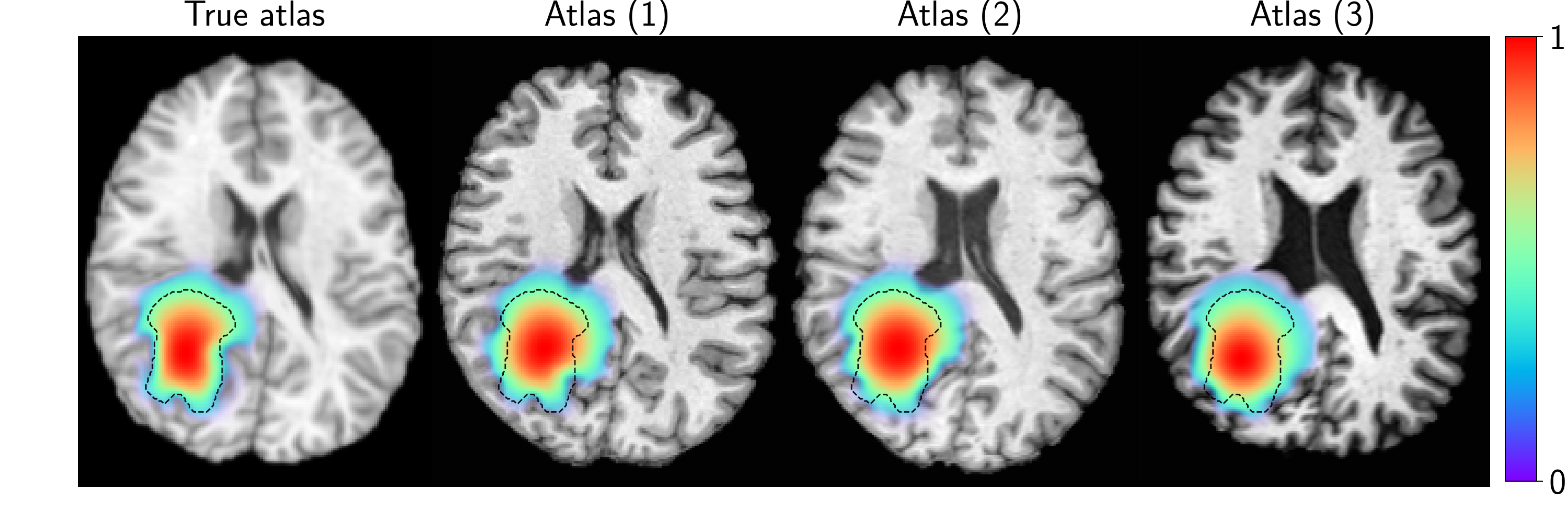}
    \caption{Reconstruction of the tumor concentration (color) using different atlases as the precancer scan. The tumor data segmentation is highlighted as a black dashed contour line.} \label{fig:syn-atlas-recon}
\end{figure}

 \medskip \noindent
\textbf{(Q3) Unknown $\vect{m}_0$, unknown $\vect{p}$: }This scenario corresponds to the actual clinical problem. For this test-case, we invoke (\textbf{S}.2) and average the results using three atlases. To reiterate the scheme, the inverted tumor ICs from  (\textbf{S}.1) are warped to each atlas (through registration) for the final inversion  (\textbf{S}.3).
We report inversion results in Tab.~\ref{tab:syn-unknown-c0-unknown-hp} and show an exemplary reconstruction of the patient using the different atlases in Fig.~\ref{fig:syn-atlas-recon}. Despite the approximation error in $\vect{m}_0$, we are still able to capture the parameters (average displacement relative errors of around 16\% and 25\% for atlases (1) and (2) respectively). Atlas (3) has poor performance because it is significantly different from the patient (for example, the ventricles are highly dissimilar). 
We also note that the error in tumor reconstruction is significantly higher, which is representative of the errors introduced due to the anatomical variations of each atlas from the patient. 
    \setlength\tabcolsep{6pt}
    \begin{table}[h!]
        \caption{Inversion results for \textbf{unknown} $\vect{m}_0$ \textbf{and} $\vect{p}$.  $e_\iota$ are the relative errors in parameter $\iota$. Atlas (*) are the different atlases used. True atlas is the ground truth $\vect{m}_0$. $u^\star$ is the ground truth displacement norm in number of voxels, $\|\vect{g}\|_2$ is the norm of the final gradient, and $\elltwoT$ is the relative error in the final tumor concentration .
             \label{tab:syn-unknown-c0-unknown-hp}}
        \centering
            \makebox[\textwidth]{\centering
                \begin{tabular}{cc|aaa|aa|cc}
                    \hline
                    ID & Test-case   & $e_\gamma$ & $e_\rho$ & $e_\kappa$  & $\|u^\star\|_\infty$ & $e_u$ & $\|\vect{g}\|_2$ & $\elltwoT$  \\
                    \hline
                    \multirow{4}{*}{\textit{TC (d)}} & True atlas &   \num{2.475000e-02} & \num{1.806420e-01} & \num{2.544840e-01} & 14.9 & \num{2.908428e-02} & \num{5.750633e-04} & \num{2.053100e-01}\\
                    & Atlas (1) &                                              \num{1.630833e-02} & \num{2.152830e-01} & \num{2.646160e-01} & 14.9 & \num{1.213299e-01} & \num{2.711289e-04} & \num{3.365050e-01}\\
                    & Atlas (2) &                                            \num{1.125500e-01} & \num{2.352750e-01} & \num{9.270400e-02} & 14.9 & \num{1.995404e-01} & \num{6.846354e-04} & \num{3.044420e-01}\\
                    & Atlas (3) &                                             \num{2.500000e-01} & \num{2.060780e-01} & \num{5.456840e-01} & 14.9 & \num{2.390996e-01} & \num{5.431947e-04} & \num{2.787730e-01}\\                      	
                    \hline
            \end{tabular}}
    \end{table}

\begin{figure}[h!]
    \centering
   \includegraphics[width=\textwidth]{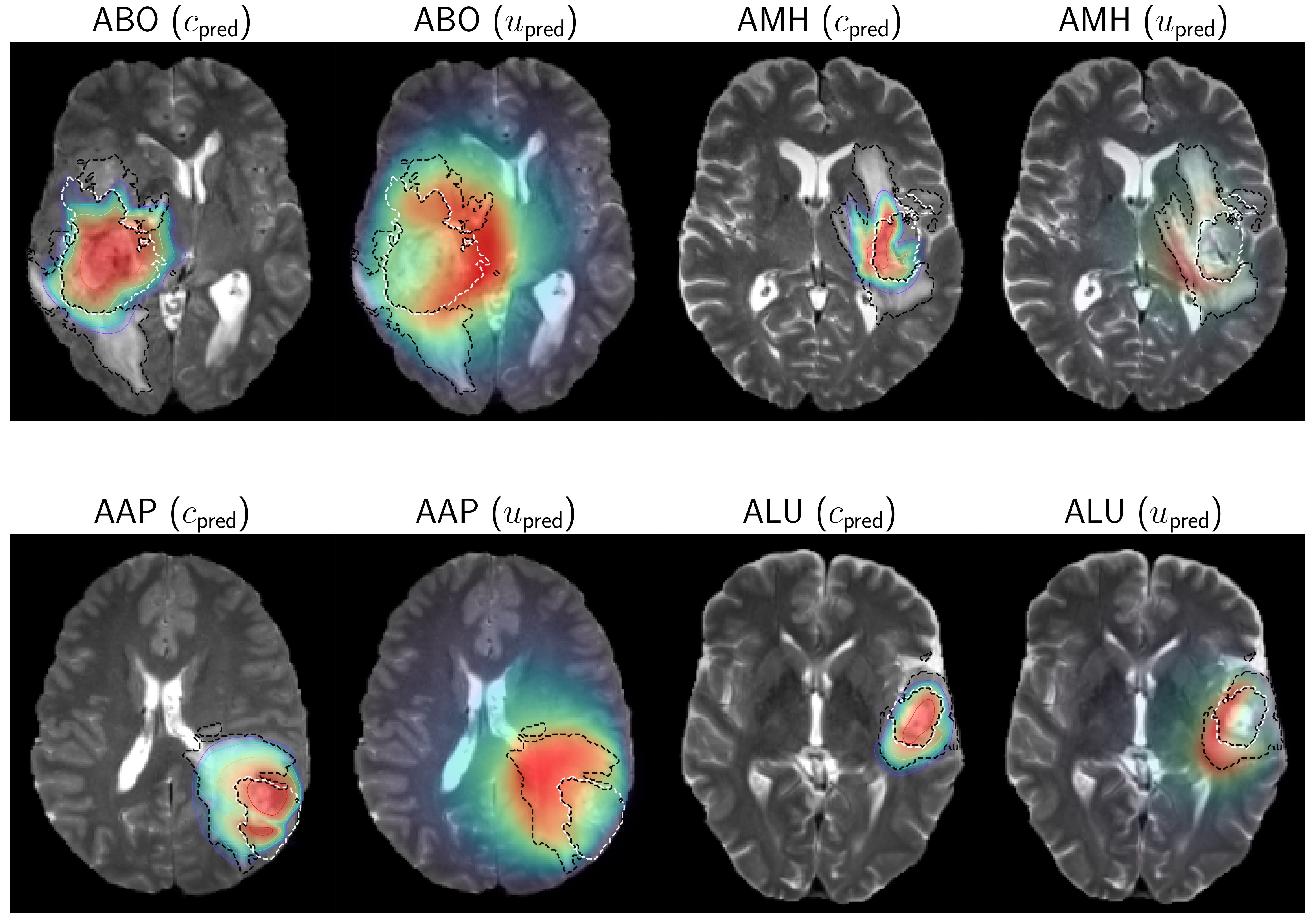}
    \caption{Predicted normalized tumor ($c_\text{pred}$) and mass effect displacement norm ($u_\text{pred}$) for each BraTS subject. The tumor data segmentation is also overlaid: the tumor core segmentation is outlined with the white dashed contour and the edema segmentation with the black dashed contour. Higher tumor concentrations ($\sim$ 1) are indicated by red and lower ones by green. Higher displacement norm values ($\sim$ 18 mm) are indicated by red and lower ones by green.\label{fig:brats}}
\end{figure}

 \medskip \noindent
\textbf{(Q4) Clinical images: }We use images from the BraTS~\cite{brats:2018} dataset. We segment the scans using a neural network based on~\cite{Gholami:2019a} and use the segmented tumor core as tumor data. We select four patients with visually different but, of course, unknown mass effect. For each patient, we invert in four different atlases used as $\vect{m}_0$.
 For each BraTS case, we report atlas-averaged model parameters.  We show each patient with the average reconstructed tumor and displacement norm in Fig.~\ref{fig:brats}. We quantify mass effect using the average maximum displacement norm, $\|u\|_\infty$. We observe that \emph{ABO} and \emph{AAP} show high mass effect ($\|u\|_\infty \sim 18~\text{mm} \pm 1$). For \emph{ALU}, the mass effect is moderate ($\|u\|_\infty \sim 8~\text{mm} \pm 2$) and is localized at the side of the brain around the tumor (see Fig.~\ref{fig:brats}); the ventricles are largely undeformed. Finally, \emph{AMH} has a mild to moderate predicted mass effect ($\|u\|_\infty \sim 5~\text{mm} \pm 3$) but exhibits the largest variation.

\section{Conclusions}
Our results are very promising. First, our solver is robust (never crashes, takes excessive iterations, or needs subject-specific hyperparameter settings). Most important,  our method  does not require any manual preprocessing and can be run in a black box fashion. Our experiments on  synthetic data with known ground truth is the first time that such a solver is verified (for example, no such verification is undertaken in GLISTR) and we demonstrated that our approximations do not introduce significant errors despite the fact that ${\vect{m}}_0$ is unknown. For clinical data, the  model errors are expected to dominate the errors from using an atlas and splitting the calibration procedure into two stages.
Second, we tested our method on a small number of clinical scans in order to test the feasibility of our method. With a very small number of calibration parameters, our solver was able to quantitatively match the observed tumor margins and qualitatively correlate with observed mass effect. We observed significant variability across subjects and small variability with respect to the choice of atlas.  Our method provides a means to quantify and localize the mass effect without relying on any assumptions on symmetry and location of the tumor. It also provides quantitative mass effect measures for comparing and stratifying subjects. The biophysical features can complement other radiogenomic features  in downstream learning tasks. Our next step is further validation using a much larger clinical dataset and the use of a large number of atlases.

\clearpage
\newpage
 \bibliographystyle{splncs04}
 \bibliography{literature}

\begin{thebibliography}{10}
\providecommand{\url}[1]{\texttt{#1}}
\providecommand{\urlprefix}{URL }
\providecommand{\doi}[1]{https://doi.org/#1}

\bibitem{Abler2019}
Abler, D., B{\"u}chler, P., Rockne, R.C.: Towards model-based characterization
  of biomechanical tumor growth phenotypes. In: Bebis, G., Benos, T., Chen, K.,
  Jahn, K., Lima, E. (eds.) Mathematical and Computational Oncology. pp.
  75--86. Springer International Publishing, Cham (2019)

\bibitem{brats:2018}
Bakas, S., Reyes, M., et~al.: Identifying the best machine learning algorithms
  for brain tumor segmentation, progression assessment, and overall survival
  prediction in the {BRATS} challenge. CoRR  \textbf{abs/1811.02629} (2018),
  \url{http://arxiv.org/abs/1811.02629}

\bibitem{clatz2005realistic}
Clatz, O., Sermesant, M., Bondiau, P.Y., Delingette, H., Warfield, S.K.,
  Malandain, G., Ayache, N.: Realistic simulation of the 3-d growth of brain
  tumors in {MR} images coupling diffusion with biomechanical deformation.
  Medical Imaging, IEEE Transactions on  \textbf{24}(10),  1334--1346 (2005)

\bibitem{Colin:2014a}
Colin, T., Iollo, A., Lagaert, J.B., Saut, O.: An inverse problem for the
  recovery of the vascularization of a tumor. Journal of Inverse and Ill-posed
  Problems  \textbf{22}(6),  759--786 (2014)

\bibitem{Gholami:2016a}
Gholami, A., Mang, A., Biros, G.: An inverse problem formulation for parameter
  estimation of a reaction-diffusion model of low grade gliomas. Journal of
  Mathematical Biology  \textbf{72}(1),  409--433 (2016)

\bibitem{Gholami2016-thesis}
Gholami, A.: Fast algorithms for biophysically-constrained inverse problems in
  medical imaging (2017), {P}h.D. dissertation thesis

\bibitem{Gholami:2019a}
Gholami, A., Subramanian, S., Shenoy, V., Himthani, N., Yue, X., Zhao, S., Jin,
  P., Biros, G., Keutzer, K.: A novel domain adaptation framework for medical
  image segmentation. In: Brainlesion: Glioma, Multiple Sclerosis, Stroke and
  Traumatic Brain Injuries. vol. LNCS 11384, pp. 289--298 (2019)

\bibitem{Gooya:2012a}
Gooya, A., Pohl, K.M., Bilello, M., Cirillo, L., Biros, G., Melhem, E.R.,
  Davatzikos, C.: {GLISTR}: {G}lioma image segmentation and registration.
  Medical Imaging, IEEE Transactions on  \textbf{31}(10),  1941--1954 (2013)

\bibitem{Hogea:2008b}
Hogea, C., Davatzikos, C., Biros, G.: An image-driven parameter estimation
  problem for a reaction-diffusion glioma growth model with mass effect. J Math
  Biol  \textbf{56},  793--825 (2008)

\bibitem{hogea2007modeling}
Hogea, C., Davatzikos, C., Biros, G.: Modeling glioma growth and mass effect in
  {3D MR} images of the brain. In: Medical Image Computing and
  Computer-Assisted Intervention--MICCAI 2007, pp. 642--650. Springer (2007)

\bibitem{hormuth2018}
Hormuth, D.A., Eldridge, S.L., Weis, J.A., Miga, M.I., Yankeelov, T.E.:
  Mechanically coupled reaction-diffusion model to predict glioma growth:
  Methodological details pp. 225--241 (2018)

\bibitem{Knopoff:2013a}
Knopoff, D.A., Fern\'andez, D.R., Torres, G.A., Turner, C.V.: Adjoint method
  for a tumor growth {PDE}-constrained optimization problem. Computers \&
  Mathematics with Applications  \textbf{66}(6),  1104--1119 (2013)

\bibitem{konukoglu2010image}
Konukoglu, E., Clatz, O., Menze, B.H., Stieltjes, B., Weber, M.A., Mandonnet,
  E., Delingette, H., Ayache, N.: Image guided personalization of
  reaction-diffusion type tumor growth models using modified anisotropic
  {Eikonal} equations. Medical Imaging, IEEE Transactions on  \textbf{29}(1),
  77--95 (2010)

\bibitem{Mang2018b}
Mang, A., Gholami, A., Davatzikos, C., Biros, G.: {CLAIRE}: A
  distributed-memory solver for constrained large deformation diffeomorphic
  image registration. SIAM Journal on Scientific Computing  \textbf{41}(5),
  C548--C584 (2019). \doi{10.1137/18M1207818}

\bibitem{mang2012biophysical}
Mang, A., Toma, A., Schuetz, T.A., Becker, S., Eckey, T., Mohr, C., Petersen,
  D., Buzug, T.M.: Biophysical modeling of brain tumor progression: from
  unconditionally stable explicit time integration to an inverse problem with
  parabolic {PDE} constraints for model calibration. Medical Physics
  \textbf{39}(7),  4444--4459 (2012)

\bibitem{murray1989mathematical}
Murray, J.D.: Mathematical biology. Springer-Verlag, New York (1989)

\bibitem{rockne-e19}
Rockne, R.C., Hawkins-Daarud, A., Swanson, K.R., Sluka, J.P., Glazier, J.A.,
  Macklin, P., Hormuth, D., Jarrett, A.M., da~Fonseca~Lima, E.A.B., Oden, J.,
  Biros, G., Yankeelov, T.E., Curtius, K., Bakir, I.A., Wodarz, D., Komarova,
  N., Aparicio, L., Bordyuh, M., Rabadan, R., Finley, S., Enderling, H.,
  Caudell, J.J., Moros, E.G., Anderson, A.R.A., Gatenby, R., Kaznatcheev, A.,
  Jeavons, P., Krishnan, N., Pelesko, J., Wadhwa, R.R., Yoon, N., Nichol, D.,
  Marusyk, A., Hinczewski, M., Scott, J.G.: The 2019 mathematical oncology
  roadmap. Physical Biology  (2019),
  \url{http://iopscience.iop.org/10.1088/1478-3975/ab1a09}

\bibitem{Scheufele:2017a}
Scheufele, K., Mang, A., Gholami, A., Davatzikos, C., Biros, G., Mehl, M.:
  Coupling brain-tumor biophysical models and diffeomorphic image registration.
  Computer Methods in Applied Mechanics and Engineering  \textbf{347},
  533--567 (2019). \doi{10.1016/j.cma.2018.12.008},
  \url{https://doi.org/10.1016/j.cma.2018.12.008}

\bibitem{scheufele2020automatic}
Scheufele, K., Subramanian, S., Biros, G.: Automatic mri-driven model
  calibration for advanced brain tumor progression analysis. arXiv pp.
  arXiv--2001 (2020)

\bibitem{Subramanian:2019a}
Subramanian, S., Gholami, A., Biros, G.: Simulation of glioblastoma growth
  using a {3D} multispecies tumor model with mass effect. Journal of
  Mathematical Biology  \textbf{79}(3),  941--967 (2019)

\bibitem{Subramanian20IP}
{Subramanian}, S., {Scheufele}, K., {Mehl}, M., {Biros}, G.: Where did the
  tumor start? an inverse solver with sparse localization for tumor growth
  models. Inverse Problems (36) (2020). \doi{10.1088/1361-6420/ab649c}

\bibitem{swanson2000quantitative}
Swanson, K., Alvord, E., Murray, J.: A quantitative model for differential
  motility of gliomas in grey and white matter. Cell Proliferation
  \textbf{33}(5),  317--330 (2000)

\bibitem{Swanson:2008:survivalResection}
Swanson, K., Rostomily, R., Alvord~Jr, E.: A mathematical modelling tool for
  predicting survival of individual patients following resection of
  glioblastoma: a proof of principle. British journal of cancer
  \textbf{98}(1), ~113 (2008)

\bibitem{swanson2002virtual}
Swanson, K.R., Alvord, E., Murray, J.: Virtual brain tumours (gliomas) enhance
  the reality of medical imaging and highlight inadequacies of current therapy.
  British Journal of Cancer  \textbf{86}(1),  14--18 (2002)

\bibitem{yankeelov-miga13}
Yankeelov, T.E., Atuegwu, N., Hormuth, D., Weis, J.A., Barnes, S.L., Miga,
  M.I., Rericha, E.C., Quaranta, V.: Clinically relevant modeling of tumor
  growth and treatment response. Science translational medicine
  \textbf{5}(187),  187ps9--187ps9 (2013)

\end{thebibliography}

\clearpage
\newpage
\section*{Supplementary materials}
\beginsupplement

 \setlength\tabcolsep{6pt}
\begin{table}[htbp!]
    \caption{Model and inversion parameter values used in our simulations. Note that the  Lam\`e coefficients $\lambda$ and $\mu$ are determined by the Young's modulus and Poisson's ratio of the tissue-type.
    }
    \centering
    \makebox[\textwidth]{\centering
        \begin{tabular}{a|c}
            \hline
            Parameter			&			value \\
            \hline
            Spatial discretization & $256^3$ \\
            Young's modulus of (GM, WM, CSF, tumor)	(Pa)&	(2100, 2100, 100, 8000)  \\
             Poisson's ratio of (GM, WM, CSF, tumor) 	&	(0.4, 0.4, 0.1, 0.45)  \\
            Gaussian width, $\sigma$ (voxels)	& 1  \\
            \hline
            Regularization parameter, $\beta$ & 1E-4 \\
            Relative gradient tolerance		& 1E-3 \\
            Relative objective function change tolerance & 1E-3 \\
            Maximum number of quasi-Newton iterations & 50 \\
            \hline
    \end{tabular}}
\end{table}

\begin{figure}[htbp!]
    \centering
    \includegraphics[width=\textwidth]{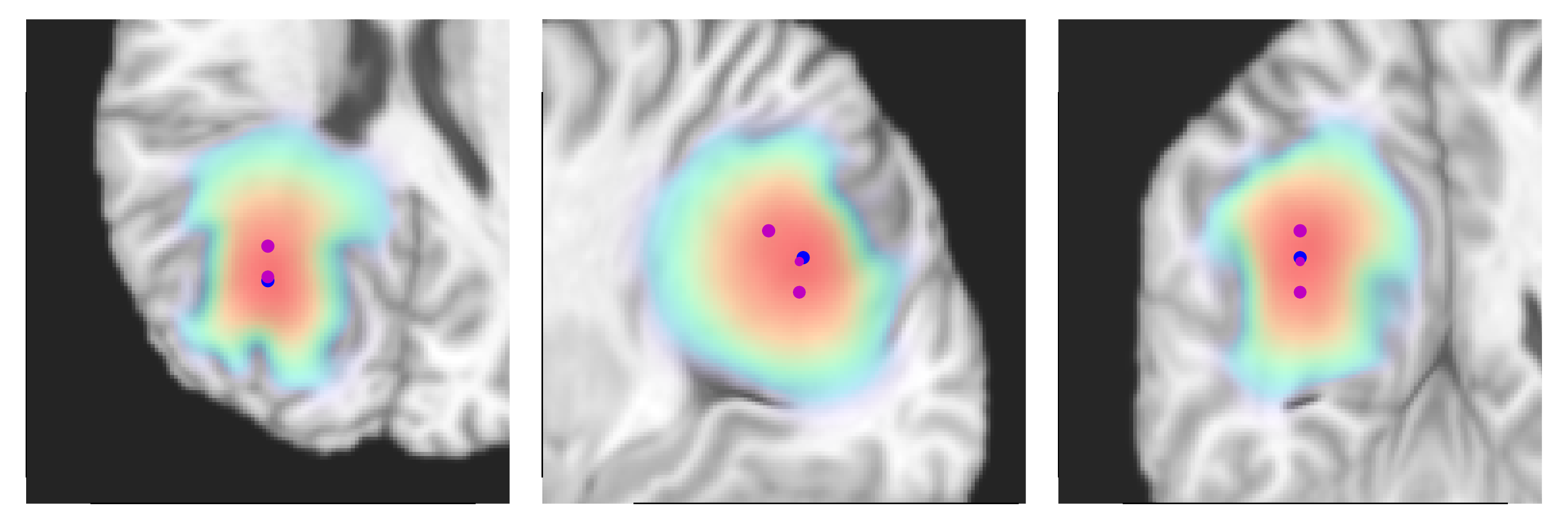}
    \caption{Estimated tumor initial conditions $\vect{p}$ reconstructed using the algorithm in (\textbf{S}.1) for test-case TC(d). The size of the markers indicate the magnitude of activation. The blue marker is the ground truth; the red markers represent the reconstruction. The tumor concentration is overlaid (red: high concentration, green: low concentration). Due to ill-posedness, the exact reconstructions cannot be recovered.} \label{fig:syn-c0-recon}
\end{figure}

 \setlength\tabcolsep{6pt}
\begin{table}[htbp!]
    \caption{Inversion results for \textbf{unknown precancer scan and tumor IC}. $u^\star$ is the ground truth displacement norm in number of voxels, $e_\iota$ is the relative error for parameter $\iota$, $e_u$ is the relative error in the two-norm of the displacement field norm, $\|\vect{g}\|_2$ is the norm of the final gradient, and $\elltwoT$ is the relative error in tumor concentration at $t = 1$. Atlas (*) are different atlases used to approximate the precancer scan; True atlas is the ground truth precancer scan. \label{tab:syn-unknown-c0-unknown-hp-supp}
    }
    \centering
    \makebox[\textwidth]{\centering
        \begin{tabular}{cc|aaa|aa|cc}
            \hline
            ID & Test-case   & $e_\gamma$ & $e_\rho$ & $e_\kappa$  & $\|u^\star\|_\infty$ & $e_u$ & $\|\vect{g}\|_2$ & $\elltwoT$  \\
                                \hline
                                \multirow{4}{*}{\textit{TC (a)}} & True atlas &  \num{6.496340e-03} & \num{1.875375e-01} & \num{2.683120e-01} & 0 & \num{6.272540e-01} & \num{2.042484e-04} & \num{2.951340e-01}\\
                                & Atlas (1) &       \num{1.302800e-01} & \num{2.845767e-01} & \num{3.324480e-01} & 0 & \num{1.120680e+01} & \num{5.208334e-04} & \num{4.764170e-01}\\
                                & Atlas (2) &      \num{5.856670e-02} & \num{2.656042e-01} & \num{3.024880e-01} & 0 & \num{5.386570e+00} & \num{1.920789e-03} & \num{4.360310e-01}\\
                                & Atlas (3) &      \num{1.881220e-01} & \num{2.743075e-01} & \num{3.440600e-01} & 0 & \num{1.687940e+01} & \num{2.129301e-03} & \num{4.844030e-01}\\
                                \hline
                                \multirow{4}{*}{\textit{TC (b)}} & True atlas &  \num{3.492500e-02} & \num{1.347417e-01} & \num{3.377520e-01} &6.1 & \num{1.471194e-02} & \num{5.690262e-04} & \num{2.233940e-01}\\
                                & Atlas (1) &    \num{1.402525e-01} & \num{2.374308e-01} & \num{1.196400e-02} & 6.1 & \num{2.399837e-01} & \num{6.999284e-04} & \num{3.853170e-01}\\
                                & Atlas (2) &     \num{1.863250e-01} & \num{2.250183e-01} & \num{1.855160e-01} & 6.1& \num{2.662218e-01} & \num{1.443246e-03} & \num{3.420640e-01}\\
                                & Atlas (3) &     \num{1.073583e+00} & \num{2.021700e-01} & \num{5.537200e-01} & 6.1& \num{8.758896e-01} & \num{4.327874e-04} & \num{3.992300e-01}\\
                                \hline
                                \multirow{4}{*}{\textit{TC (c)}} & True atlas & \num{1.031250e-02} & \num{1.795880e-01} & \num{2.466480e-01} & 10 & \num{6.277826e-02} & \num{3.659728e-04} & \num{1.776210e-01} \\
                                & Atlas (1) &   \num{5.838750e-03} & \num{2.213250e-01} & \num{3.477000e-01} & 10  & \num{1.056786e-01} & \num{9.176429e-05} & \num{3.181160e-01}\\
                                & Atlas (2) &   \num{2.081913e-01} & \num{2.166190e-01} & \num{2.005360e-01} & 10 & \num{2.749868e-01} & \num{1.498397e-04} & \num{2.872030e-01}\\
                                & Atlas (3) &   \num{6.081125e-01} & \num{1.805890e-01} & \num{5.630280e-01} & 10  & \num{4.901243e-01} & \num{7.106825e-04} & \num{3.564560e-01}\\
            \hline
            \multirow{4}{*}{\textit{TC (d)}} & True atlas &   \num{2.475000e-02} & \num{1.806420e-01} & \num{2.544840e-01} &14.9 & \num{2.908428e-02} & \num{5.750633e-04} & \num{2.053100e-01}\\
            & Atlas (1) &                                              \num{1.630833e-02} & \num{2.152830e-01} & \num{2.646160e-01} &14.9 & \num{1.213299e-01} & \num{2.711289e-04} & \num{3.365050e-01}\\
            & Atlas (2) &                                            \num{1.125500e-01} & \num{2.352750e-01} & \num{9.270400e-02} &14.9 & \num{1.995404e-01} & \num{6.846354e-04} & \num{3.044420e-01}\\
            & Atlas (3) &                                             \num{2.500000e-01} & \num{2.060780e-01} & \num{5.456840e-01} &14.9 & \num{2.390996e-01} & \num{5.431947e-04} & \num{2.787730e-01}\\                      	
            \hline
    \end{tabular}}
\end{table}

\setlength\tabcolsep{3pt}
\begin{table}[htbp!]
    \caption{Inversion results for \textbf{clinical images} using statistics from four atlases. We report average parameter values for $(\gamma, \rho, \kappa, \|u\|_\infty)$ along with the  average relative error in tumor reconstruction ($\elltwoT$) at observation points. We also report the standard deviation for each metric. \label{tab:brats}
    }
    \centering
    \makebox[\textwidth]{\centering
        \begin{tabular}{c|aaa|a|c}
            \hline
            Patient-ID   & $\gamma$ & $\rho$ & $\kappa$  & $\|u\|_\infty$ (in voxels)  & $\elltwoT$  \\
            \hline
            \textit{ABO} & \num{8.781998e-01} $\pm$ \num{3.973393e-02} & \num{1.374553e+01} $\pm$ \num{6.454700e-01} & \num{7.355470e-03} $\pm$ \num{1.766605e-03} & 20.9 $\pm$ 1.11 &  \num{4.370733e-01} $\pm$ \num{2.253797e-02} \\
            \textit{AMH} & \num{4.438890e-01} $\pm$ \num{3.331645e-01} & \num{1.455325e+01} $\pm$ \num{5.306997e-01} & \num{9.984125e-03} $\pm$ \num{5.003036e-03} & 5.55 $\pm$ 3.95 &  \num{6.477607e-01} $\pm$ \num{3.010800e-02} \\ 
            \textit{AAP} & \num{1.403883e+00} $\pm$ \num{8.846973e-02} & \num{1.374862e+01} $\pm$ \num{8.629452e-01} & \num{5.000888e-03} $\pm$ \num{1.332167e-06} & 21.2 $\pm$ 0.94 &  \num{5.756205e-01} $\pm$ \num{1.266747e-02} \\
            \textit{ALU} & \num{1.081802e+00} $\pm$ \num{2.368687e-01} & \num{9.111317e+00} $\pm$ \num{1.807181e-01} & \num{5.000000e-03} $\pm$ \num{0.000000e+00} & 9.23 $\pm$ 2.08 &  \num{4.424365e-01} $\pm$ \num{4.359449e-02} \\
            \hline
    \end{tabular}}
\end{table}
\end{document}